\documentclass[a4paper,10pt]{article}
\usepackage[all]{}
\usepackage{graphicx}
\usepackage{epsfig}

\begin{document}

\title{A useful correspondence between fluid convection and ecosystem operation} 

\maketitle

\begin{center}
{\bf Stijn Bruers}\footnote{email: stijn.bruers@fys.kuleuven.be}\\
Instituut voor Theoretische Fysica, \\
Katholieke Universiteit Leuven,\\
Celestijnenlaan 200D, B-3001 Leuven, Belgium
\vspace{15pt}

{\bf Filip Meysman}\footnote{email: f.meysman@nioo.knaw.nl}\\
Centre for Estuarine
and Marine Ecology,\\
Netherlands Institute of Ecology (NIOO-KNAW),\\
Korringaweg 7, 4401 NT Yerseke, The Netherlands\\
\end{center}

\begin{abstract}
Both ecological systems and convective fluid systems are examples of open systems which operate
far-from-equi\-li\-bri\-um. This article demonstrates that there is a correspondence between a resource-consumer chemostat ecosystem and the Rayleigh-B\'enard (RB) convective fluid system. The Lorenz dynamics of the RB system can be translated into an ecosystem dynamics. Not only is there a correspondence between the dynamical equations, also the physical interpretations show interesting analogies. By using this
fluid-ecosystem analogy, we are able to derive the correct value of
the size of convection rolls by competitive fitness arguments
borrowed from ecology. We finally conjecture that the Lorenz dynamics can be extended to describe more complex convection patterns that resemble ecological predation.
\end{abstract}

\noindent\emph{Keywords:} Rayleigh-B\'enard system; Lorenz model; resource-consumer chemostat; ecosystem metabolism; thermodynamics

\section{Introduction}

In a Rayleigh-B\'enard experiment, a horizontal viscous fluid layer is heated from below. When the temperature difference between upper and lower sides is small,
heat transfer solely occurs through thermal conduction. Yet once
beyond a critical temperature difference, a regular pattern of
convection cells or rolls emerges (B\'enard, 1901). This sudden shift
from conduction to convection is referred to as the Rayleigh-B\'enard (RB)
instability, and is often quoted as an archetypal example of
self-organization in non-equilibrium systems (Nicolis and Prigogine, 1989; Prigogine, 1967).

Intuitively, it makes sense to try to apply the concept of
self-organization in non-equilibrium systems to ecologicaly systems,
as there are similarities between ecological and physical systems.
Like the Rayleigh-B\'enard set-up, ecological systems
are open systems that receive a throughput of energy and/or mass via
coupling to two environments (Morowitz, 1968; Schr{\"o}dinger, 1944). These two environments are typically large reservoirs and they drive the system from equilibrium. Consider the example of a laboratory chemostat ecosystem (Smith and Waltman, 1995). This is a prime example of a chemotrophic ecosystem whereby a resource of energetic high quality chemical substrate is pumped from a reservoir into the system. In the ecosystem this resource is degraded into low quality waste products which are emitted to the waste reservoir. When there is low feeding of resource, no biota can survive, and the resource is degraded by abiotic processes only. But when the feeding is above a critical threshold, biota can survive by consuming the resource. There is a sudden shift from a lifeless to a living state\footnote{Strictly speaking, it is rather a distinction between guaranteed extinction and survival. We study what will happen with an organism which is released in the ecosystem. Our results should not be interpreted as the solution for the origin of life.}. In other words, the energetic quality difference
between incoming and outgoing chemical substrates is exploited by various abiotic and biotic processes. The latter biotic processes contain the biomass synthesis and turnover of consumer micro-organisms feeding on the resource. 

So it is tempting to look for a deeper connection. Can one compare biological processing with convection? Both mechanisms involve self-organizing structures, biological cells or convection cells, that can only survive after a critical threshold. Both energetic pathways, biotic resource conversion and thermal convection, degrade energy from high quality to low quality form. And these energetic pathways are additional to the abiotic conversion or thermal conduction processes of the background.

Here, our ambition is to
examine the link between ecological processes and convective fluid motions in a quantitative way. The first part of this article contains a highly intriguing result: The mathematical expressions of the resource-consumer chemostat ecosystem dynamics are exactly the same as the dynamics that describes the basics of the Rayleigh-B\'enard system. Furthermore, not only are the mathematical equations identical, also the physical/ecological interpretations give appealing results. Particularly, by looking at the energetic pathways of the ecosystem, the ecological quantities can be mapped to the quantities used in the fluid system and vice versa.

The second part tries to extend the correspondence between fluid convection and ecosystem functioning to include new processes. We will study two extensions. First, one can look at ecological competition and translate the notion of competitive fitness to the fluid system. The convection cells are in 'Darwinian competition' with each other and the fittest ones will survive. One can generalize the Lorenz model to include this fluid competition. As the size of a convection cell will depend on the fitness measure, we will demonstrate that the mathematical identity of the ecological and the fluid dynamics predicts the experimentally correct size of the cells at the onset of convection. Second, one can look at ecological predation. Translating this notion to the fluid system leads to a new conjecture to extend the Lorenz model in order to describe more complex convection patterns. These new patterns only appear when the system is driven beyond a second critical value for the temperature gradient. The 'predatory behavior' in the convective fluid system leads us to a conjecture which we will not prove, but will be successfully tested by looking at the energy dissipation.

\section{The Ray\-leigh-B\'enard convection system}

Let us start by deriving the dynamics that describes the Rayleigh-B\'enard (RB) convective fluid system, named after B\'enard (1901) and Ray\-leigh (1916) who were the first to study this system experimentally and theoretically. 
A full mathematical treatment of thermal convection requires the
combined solution of the heat transport, Navier-Stokes and
incompressibility equations, resulting in a set of five coupled non-linear
partial differential equations (Chandrasekhar,
1961; Ray\-leigh, 1916). 

\begin{figure}[!ht]
\centering
\includegraphics[scale=0.5]{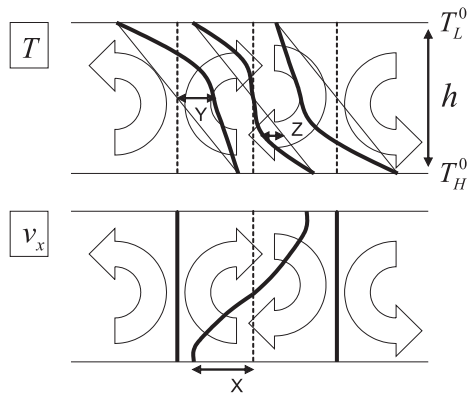}
\caption{The profiles and the variables $X$, $Y$ and $Z$. The temperature ($T$) and horizontal velocity ($v_x$)
profiles  at three vertical sections (dashed lines) are shown. These
vertical sections are parallel with the axes of the convection
rolls, where the fluid is moving up, moving horizontal or moving
down. The thin linear profiles correspond with the conduction state,
the thick profiles with the convection state. As indicated, $Y$ and
$Z$ are temperature deviations and $X$ is the velocity at the bottom of
a roll.} \label{YZandXfig}
\end{figure}

Rather than solving this full set, we employ the
approximation adopted by Lorenz (1963), which became famous as it gave an impulse to the development of chaos theory. The model describes the lowest modes of an expansion of the temperature and velocity fields for a RB system with free-free boundary conditions (see e.g. Getling 1998). In appendix \ref{appendix XYZ}, the derivation of the Lorenz system is given in a way that will suit our further discussion. A non-linear set of three ordinary differential equations is obtained, with three variables (see fig. \ref{YZandXfig}): $X$ measures the rotational rate of the rolls and represents the maximal velocity at the bottom of the rolls. $Y$ and $Z$ are temperature deviations, where the linear profile of the conduction state is taken as a reference. 

With these three variables, the XYZ Lorenz system is rich enough to describe the Rayleigh-B\'enard instability, the sudden shift
from conduction to convection. But there is an even simpler model, the XZ system with only two variables, that is rich enough as well. It is this XZ model that allows us to make the correspondence. Roughly speaking, we will perform an averaging over the horizontal directions, such that only the average vertical profile remains. As $Y$ is the temperature deviation in horizontal direction, it is this variable that will disappear after the averaging. Specifically, this is done by making the pseudo steady state
assumption $dY/dt=0$ for the variable $Y$. The latter becomes a constant and the dynamics turns into:
\begin{eqnarray}
\frac{d X}{d t}&=&\left( \frac{a^2h^2g\alpha}{(a^2+1)^2\pi^2\chi}-\frac{(a^2+1)\pi^2\nu}{h^2}\right) X -\frac{2a^2hg\alpha}{ (a^2+1)^2\pi\chi}XZ, \label{dyn X}\\
\frac{d Z}{d t}&=&\frac{a^2 \pi \beta}{2(a^2+1)h\chi}X^2-\frac{a^2\pi^2}{(a^2+1)h^2\chi}X^2Z-\frac{4\pi^2 \chi}{h^2}Z, \label{dyn Z}
\end{eqnarray}
with $h$ the height of the fluid layer, $a$ a geometric factor such that $h/a$ is the width of the straight convection rolls, $\alpha$ the thermal
expansion coefficient, $g$ the gravitational acceleration, $\chi$
the heat conduction coefficient, $\nu$ the kinematic viscosity and 
\begin{eqnarray}
\beta=\frac{T_H^0-T_L^0}{h}
\end{eqnarray}
the temperature gradient. This important quantity is the thermodynamic gradient that drives the system out of equilibrium. $T_H$ is the high temperature of the heat reservoir below the fluid layer and $T_L$ is the low temperature of the heat reservoir above the layer.

\begin{figure}[!ht]
\centering
\includegraphics[scale=0.5]{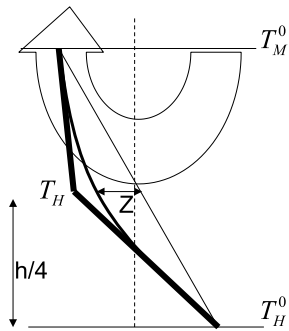}
\caption{The definition of $T_H$. The lower half of the fluid layer is shown,
with the vertical (horizontally averaged) temperature profiles in
the conduction state (thin line), the convection state (thicker
line) and the 'linearized' convection profile (thickest line). The
variables $Z$ and $T_H$ are at height $h/4$. } \label{THfig}
\end{figure}

For further reference, we will also need measures for the temperatures at the middle and the lower halve of the fluid layer. Define
\begin{eqnarray}
T_M^0 \equiv \frac{T_H^0+T_L^0}{2}
\end{eqnarray}
as the horizontally average temperature at height $h/2$, and
\begin{eqnarray}
T_H \equiv \frac{T_H^0+T_M^0-\pi Z}{2}. \label{T_H}
\end{eqnarray}
Fig. \ref{THfig} shows the interpretation of $T_H$ as a temperature measure for a linearized temperature profile in the lower half of the fluid layer. (Due to symmetry in the
approximation leading to the Lorenz system, we will not have to
include the upper half of the fluid layer.)

\section{The resource-consumer ecosystem}\label{chemotrophic ecosystem}

Next, we discuss the ecosystem model, which is in essence a simple chemotrophic resource-consumer food
web model, one of the mainstay models of ecology (e.g.
Yodzis and Innes, 1992). Consumer organisms are feeding on some
food resource (R), which is partly converted to consumer biomass (C)
and partly to waste product (W). For reference, one can think of a chemostat set-up where a chemical reactor tank contains a monoculture of micro-organisms that are feeding on a chemical substrate like methane or glucose, while respiring $CO_2$.

\begin{figure}[!ht]
\centering
\includegraphics[scale=0.5]{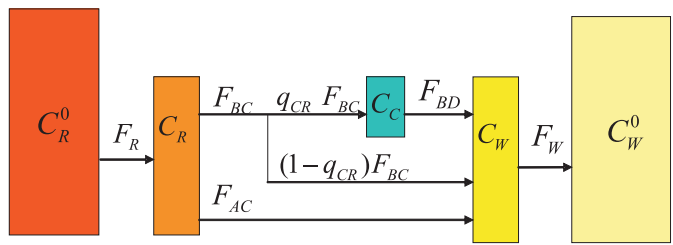}
\caption{The ecosystem flow chart. The resource-consumer-waste ecosystem coupled with the two environments. The different fluxes are discussed in appendix \ref{appendix RCW}. The color denotes the 'energetic quality' of the substances, from high (red) to low (light yellow).}
\label{RCWmodelfig}
\end{figure}

Figure \ref{RCWmodelfig} gives a schematic overview of the ecosystem coupled with the two environments (denoted with the superscripts $0$). There are two environmental compartments, the resource at constant concentration $C_R^0$ and the waste at constant concentration $C_W^0$) and three ecosystem compartments, with variable concentrations $C_R$, $C_C$ and $C_W$ for the resource, the consumer biomass and the waste respectively. In appendix \ref{appendix RCW}, the complete dynamics of the re\-sour\-ce-con\-su\-mer-wa\-ste (RCW) ecosystem is given, explaining the fluxes $F$ between the compartments. 

However, as we will see, the correspondence only works in a limiting case, whereby roughly speaking we will average over the waste concentrations of the system and the environment. Specifically, this can be done by taking a very small relaxation time for the exchange of the waste between the ecosystem and the reservoir. This means that by studying the ecosystem at longer time scales than this relaxation time, the dynamics for W is forced to be in a pseudo steady state condition. Hence, W is no longer a variable and we end up with the resource-consumer (RC) model, with two dynamical equations for two variables:
\begin{eqnarray}
\frac{d}{dt}C_R &=& \alpha_R(C_R^0 - C_R) -(\kappa_{AC}+g_{CR}
C_C)(C_R-C_W^0/K_{eq}), \label{dyn C_R} \\
 \frac{d}{dt}C_C &=&
q_{CR} g_{CR} (C_R-C_W^0/K_{eq}) C_C - d_C C_C, \label{dyn C_C}
\end{eqnarray}
with, $\alpha_R$ the resource exchange rate parameter, $\kappa_{AC}$ the abiotic conversion (from R to W) rate parameter, $g_{CR}$ the consumer growth rate parameter, $q_{CR}$ the yield factor for the consumer growth, $d_C$ the consumer decay (biomass turnover) rate parameter, and $K_{eq}$ the equilibrium constant for the chemical reaction (oxidation) from R to W which always slowly proceeds at the background. 

This is the well-known chemostat dynamics (Smith and Waltman, 1995), which is extended in two ways: First, abiotic conversion is included in terms of chemical oxidation with parameters $\kappa_{AC}$ and $C_W^0/K_{eq}$. Second, instead of the classical dependence of the growth on the resource $C_R$, the growth is now made dependent on $C_R-C_W^0/K_{eq}$. This is done for thermodynamic consistency: at chemical equilibrium, biomass synthesis should also cease. 
\newline

\begin{table}[!ht]
\begin{center}
\begin{tabular}{|c||c|c|}
  \hline
  & ecosystem & fluid system\\
  \hline
  \hline
  unstructured \& & abiotic \& & conductive \& \\
  structured & biotic & convective\\
  gradient & ecosystem & heat \\
  degradation & metabolism & transport\\
  \hline
  structures & biological organisms & convection patterns \\
  \hline
  model & RC & XZ \\
  \hline
\end{tabular}
\end{center}
\caption{Two corresponding models with analogous mechanisms for gradient degradation}\label{table gradient degradation}
\end{table}

Let us summarize. Table \ref{table gradient degradation} shows the observation that there are two systems with analogous mechanisms for the degradation of a gradient, i.e. the transformation of high quality energy to low quality energy. The unstructured processes are the ground level mechanisms: abiotic conversion from resource to waste or thermal conduction from high temperature to low temperature. But above a certain critical threshold, a self-organization mechanism adds second level processes: biotic conversion or thermal convection.  

To study these systems, we introduced two models, each with three variables: the XYZ Lorenz model (one velocity $X$ and two temperatures $Y$ and $Z$) and the RCW ecosystem model (one biotic consumer $C$ and two abiotic molecules $R$ and $W$). These systems have different behavior, as the XYZ model has chaotic solutions whereas they are absent in the RCW model. However, there is a hidden correspondence which we will clarify in the next section. We have to make a pseudo steady state condition (an averaging) of the 'abiotic' variables $Y$ and $W$, leading to the XZ model (\ref{dyn X}-\ref{dyn Z}) and the RC model (\ref{dyn C_R}-\ref{dyn C_C}). These models have only two variables and hence they are the most simple models to study a non-trivial behavior, the transition from an 'abiotic' to a 'biotic' state. The XZ system does not have chaotic solutions anymore, so it is possible that it is mathematically equivalent with the RC model. The trick is to rewrite the variables and the parameters to demonstrate this equivalence. To give a first hint, the basic observation is that the variables should be related as
\begin{eqnarray}
C_R-\frac{C_W^0}{K_{eq}} &\leftrightarrow& \frac{h\beta}{4}-\frac{\pi}{2}Z, \label{C_R Z}\\
C_C &\leftrightarrow& \frac{X^2}{g\alpha h^3}. \label{C_C=X^2}
\end{eqnarray}
Note that the quantities on the right hand side have dimensions of temperature.
In the next section, we will also relate the parameters and discuss the physical interpretations of this correspondence

\section{The correspondence}

So it is time to write the dictionnary of the correspondence. Our final result is shown in table \ref{table correspondence} at the end of this article. In order to reach our goal, we need to be able to consistently translate quantities from one system to the other. The redefinitions explained below enable us to write the simplified Lorenz dynamics as the ecosystem dynamics.

First we will state the relation between the basic quantities, the concentrations and the temperatures, which is simply:
\begin{eqnarray}
C_R^0 &\leftrightarrow& T_H^0, \label{C_R^0=T_H^0}\\
C_R &\leftrightarrow&  T_H, \label{C_R T_H}\\
\frac{C_W^0}{K_{eq}} &\leftrightarrow& T_M^0.
\end{eqnarray}
These were derived by using (\ref{C_R Z}) and the interpretation of $T_H$ (\ref{T_H}). It explains why we can roughly interpret the resource as the heat energy. \\
The consumer concentration is given by (\ref{C_C=X^2}).
As $X$ is a velocity measure, $X^2$ is a measure for the kinetic energy of the convection rolls. This kinetic energy is consuming the heat energy resource.\\
The yield and consumer growth parameters are written as
\begin{eqnarray}
q_{CR}&\leftrightarrow& \frac{8}{\pi^4(a^2+1)},\\
g_{CR}&\leftrightarrow& \frac{g\alpha h a^2 \pi^2}{(a^2+1)\chi}.
\end{eqnarray}
As the gravitational field is causing the buoyancy force, this explains why $g$ appears in $q_{CR}$. Furthermore, these parameters depend on geometric factors, especially $h/a$, the width of a convection roll. The importance of this dependence will be shown later.\\
The abiotic exchange and abiotic conversion parameters are
\begin{eqnarray}
\alpha_R&=&\kappa_{AC}\leftrightarrow\frac{2\pi^2 \chi }{h^2}.
\label{alphaR=chi}
\end{eqnarray}
As these parameters are conduction coefficients, it is logical that they depend on the heat conduction coefficient $\chi$. In order that the analogy works, our ecosystems should have equal exchange and abiotic conversion parameters\footnote{
The reason is that we identified (\ref{C_R T_H}), and the distance between the lower side and height $h/4$ equals the distance from this height to the middle of the fluid layer. There is a possibility to have a more general correspondence, with $\alpha_R\neq \kappa_{AC}$, but then we will loose the relation (\ref{C_R T_H}).}.\\
The final parameter is the biomass decay rate
\begin{eqnarray}
d_C&\leftrightarrow& \frac{2 (a^2+1)\pi^2 \nu}{h^2}. \label{dC=nu}
\end{eqnarray}
This explains why this decay is a kind of friction term. As
mortality and viscous friction destroy the biological or convective cells, a
continuous feeding on the resource is required in order that these
structures can survive.

Having discussed the relations between variables and parameters of
both systems, one can take a look at other ideas and concepts of one
system and translate it to the other. A quantity that will become useful later is the thermodynamic gradient that measures how far the system is out of equilibrium. It is given by the difference in energetic 'quality' of the two reservoirs.
\begin{eqnarray}
\Delta^0 \equiv C_R^0-\frac{C_W^0}{K_{eq}} \leftrightarrow \frac{h\beta}{2}=(T_H^0-T_M^0). \label{Delta^0=hbeta}
\end{eqnarray}
The latter relation can be turned into a dimensionless measure, which is the well known Rayleigh number $Ra$ in fluid systems. Another important dimensionless fluid quantity is the Prandtl number $Pr$. We can now see that they can be casted into their ecological analogs:
\begin{eqnarray}
Ra &\equiv& \frac{g\alpha h^3 (T_H^0-T_L^0)}{\nu\chi} \leftrightarrow G\frac{q_{CR}g_{CR}\Delta^0}{d_C}, \label{Ra}\\
Pr &\equiv& \frac{\nu}{\chi} \leftrightarrow
H\frac{d_C}{\alpha_R+\kappa_{AC}}. \label{Prandtl}
\end{eqnarray}
The geometric factors
\begin{eqnarray}
G&=&\frac{(a^2+1)^3\pi^4}{ 2a^2}, \label{G}\\
H&=&\frac{a^2+1}{2}
\end{eqnarray}
will become important later on.

This is the first part of our dictionarry. In the next section we will delve deeper into the physical analogies between both systems. In particular, we will look at the energy dissipation along the different energetic pathways.

\section{Energy flows along energetic pathways}

Our next challenge is to see whether the correspondence also works for the energy flows along the different pathways. Are the heat transport and the ecosystem metabolism connected? This question is not trivial, because
even though the dynamical equations look the same, a priori it is
not obvious that the thermodynamical expression for the heat
transport is exactly the same term in the dynamical equations which
corresponds with the ecosystem metabolism rate. Schneider and Kay (1994, fig 2a) used experimental data sets for the RB system to plot the total steady state vertical heat transport per unit horizontal
area $W^*$. (Steady states are denoted with a superscript $*$.) Our approach now allows us to write down a
simple analytical expression for the heat transport in the steady
state, because it will be shown to be related with the total ecosystem
metabolism (the total rate of waste production, see (\ref{dyn C_W}))
\begin{eqnarray}
F_{EM}\equiv(\kappa_{AC} +(1-q_{CR}) g_{CR} C_C)(C_R-C_W^0/K_{eq})+d_C C_C. \label{F ecosystem metabolism}
\end{eqnarray}
Our result will suit well the behavior as seen in the plot derived by Schneider and Kay\footnote{Schneider and Kay (1994) described a fluid layer with rigid-rigid boundary conditions. Therefore, our results can only be compared qualitatively, as our XZ model only works for systems with free-free boundary conditions.}.

To calculate $W^*$, observe that there is no energy accumulation in
the fluid, and hence this heat transport is the same at every
height. Therefore it equals the transport at height $z=0$. At the
bottom layer, the vertical fluid motion is zero, as is seen in the chosen boundary condition (\ref{v_z(z=0)}). Hence, at the bottom layer there is no vertical heat transport by fluid motion. The heat
transport is given by the temperature gradient only, as for the
conduction state. Taking a horizontal average, the $Y$-term in the expansion (\ref{Temperature expansion}) drops out, leaving only the $Z$-term. This gives:
\begin{eqnarray}
(\rho_0 c_V)^{-1} W^*&=&-\chi\frac{\partial T^*}{\partial
z}|_{z=0}\\
&=&\chi(\beta+\frac{2\pi
Z^*}{h})\\
&=&\frac{2h}{\pi^2}\alpha_R(C_R^0-C_R^*).
\label{energydissipation}
\end{eqnarray}
($\rho_0$ is the reference density and $c_p$ is the heat capacity.) The latter expression gives the steady state resource exchange $F_R^*$, which equals $F_{EM}^*$ (this is easily seen because there is no accumulation of ecosystem resource or biomass, and hence the net resourche exchange should equal the total conversion from resource to waste).

In order to study the behavior of the ecosystem metabolism $F_{EM}^*$ under different gradients, we need to solve the dynamics for the steady states. Scanning $\Delta^0$ from zero to infinity, there is a critical value
given by the bifurcation point
\begin{eqnarray}
\Delta_c^0=\frac{(\alpha_R + \kappa_{AC}) d_C}{\alpha_R q_{CR}
g_{CR}} \label{C_Rcrit}
\end{eqnarray}
For a value of $\Delta^0\leq \Delta_c^0$, we have only one stable steady state
that is physically realistic (no negative concentrations)
\begin{eqnarray}
C_R^* &=& \frac{\alpha_R \Delta^0}{\alpha_R +\kappa_{AC}}+\frac{C_W^0}{K_{eq}}, \label{C_R^*1}\\
C_C^* &=& 0.
\end{eqnarray}
Within this region, a stable population of consumers cannot be
formed, and hence, only abiotic degradation takes place. However, if
the resource input increases so that $\Delta^0\geq \Delta_c^0$,
there is the possibility for the consumers to survive at a non-zero
concentration. The above state becomes unstable, and the new stable
solution becomes
\begin{eqnarray}
C_R^* &=& \frac{d_C}{q_{CR} g_{CR}}+\frac{C_W^0}{K_{eq}}, \\
C_C^* &=& \frac{q_{CR}g_{CR}\alpha_R \Delta^0
-(\alpha_R+\kappa_{AC})d_C}{d_Cg_{CR}}. \label{C_C^*2}
\end{eqnarray}

\begin{figure}[!ht]
\centering
\resizebox{0.48\textwidth}{!}{\includegraphics{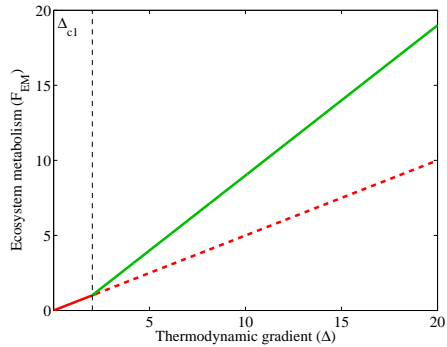}}
\caption{The total steady state ecosystem metabolism $F_{EM}^*$ in function of the driving force $\Delta^0$,
for specific parameter values. Red color denotes abiotic conversion
only, the green line corresponds with biotic consumption. The dashed
line corresponds with unstable states.} \label{ratesconsumerfig}
\end{figure}

Using these solutions, we can plot the steady state ecosystem metabolism $F_{EM}^*$ as a function of the thermodynamic gradient $\Delta^0$, Figure \ref{ratesconsumerfig}. The (qualitative) similarity with the Figure 2a in Schneider and Kay (1994) is obvious. The steady states which have only abiotic conversion are located at the so called thermodynamic branch, because this branch contains thermodynamic equilibrium at zero gradient ($F^*_{EM}=0$ at $\Delta^0=0$). In the RB system, these states correspond with thermal conduction. But above the bifurcation point, there is an exchange of stability: the thermodynamic branch states become unstable and new stable states arise. These are located at the so called dissipative branch, and they contain both abiotic and biotic degradation of resource. Once beyond the bifurcation, a viable consumer population can be established. Translated to the RB system, both conductive and convective heat transport processes appear and a viable 'kinetic energy population' is established.

\begin{figure}[!ht]
\centering
\includegraphics[scale=0.5]{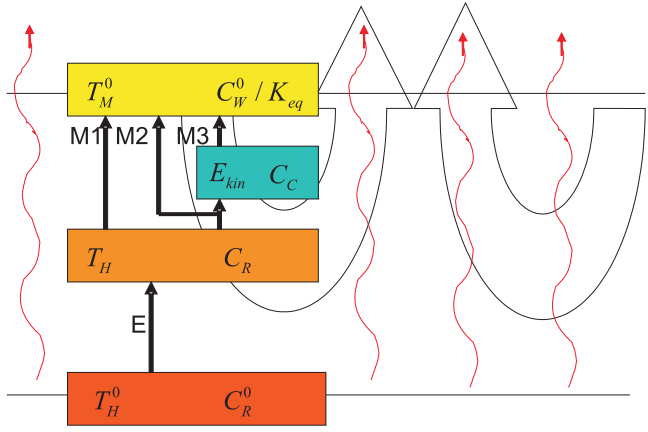}
\caption{The correspondence between temperatures and concentrations, as well
as the energetic pathways. The figure shows the lower half of the convective fluid layer, as in Fig. \ref{THfig}, with vertical energy transport. The boxes represent the heat energy compartments at three different heights, plus the kinetic energy compartment. In this way, the Rayleigh-B\'enard system is presented in terms of a simple resource-consumer food web. See text for more explanations.}
\label{energeticpathwaysfig}
\end{figure}

The above discussion shows the exact correspondence between two terms in the dynamics that describe the energy dissipation: the heat transport and the ecosystem metabolism. However, the argument was restricted to the steady state behavior. We will now give some other arguments to demonstrate that there is not only a formal \emph{mathematical} equivalence of the RC and the XZ models, but that the terms in the dynamical equations correspond also \emph{physically} with the different energetic pathways, Fig. \ref{energeticpathwaysfig} (compare with Fig. \ref{RCWmodelfig}). This correspondence of the energetic pathways of both systems is also valid in the transient states.

First let us look at the exchange with the external reservoir (E).
The fluid has a heat exchange with the heat reservoir at constant
temperature $T_H^0$. This exchange is due to heat conduction with
coefficient $\chi$. The ecosystem has the same functioning: The
variable $C_R$ is in contact with the constant $C_R^0$ with exchange
rate $\alpha_R$, explaining the relation (\ref{alphaR=chi}) and
(\ref{energydissipation}).

Next, let us focus on the energetic pathways within the system. In our
resource-consumer ecosystem, we have seen that there are basically three meta\-bo\-lic
pathways for the consumption of the resource $C_R$ (see Table \ref{table ecosystemtransformations} in appendix \ref{appendix RCW}). These are the
three arrows arriving at the waste compartment $C_W^0$ in the
figure.

Also our fluid system has three equivalent heat transport and energy
transformation pathways (see e.g. (\ref{F ecosystem metabolism})) :
\begin{itemize}
\item M1: There is heat transport by conduction which is qualitatively given by $W_{cond} \propto \chi(T_H-T_M^0)/(h/4)$, and this is indeed proportional with the abiotic conversion
\begin{eqnarray}
F_{AC}=\kappa_{AC}(C_R-C_W^0/K_{eq}).
\end{eqnarray}
\item M2: There is direct heat transport by convection, i.e. heat energy from the lower reservoir is actively transported to the upper reservoir, without being turned in kinetic energy. This is a loss term for the transformation of heat energy to kinetic energy. It is easily seen that this term is proportional with the consumer consumption (which is coupled with the consumer growth)
\begin{eqnarray}
(1-q_{CR})F_{BC}=(1-q_{CR})g_{CR}(C_R-C_W^0/K_{eq})C_C,
\end{eqnarray} 
because $C_C$ is the kinetic energy. Using the dictionary, one can translate this expression into an analytical expression for the direct heat transport by convection $W_{dir,conv}$.
\item M3: There is heat production due to viscous dissipation of kinetic energy. This extra heat produced is also finally released in the cold temperature reservoir. It is the indirect heat transport by convection, as the heat energy is first turned into kinetic energy, and eventually released again as heat energy. In Kreuzer (1981), a derivation is given for this transformation rate of kinetic energy into heat energy:
\begin{eqnarray}
W_{indir,conv}=\frac{\nu}{2}\sum_{\{i,j\}=\{x,z\}}\left(
\frac{\partial v_i}{\partial r_j}+\frac{\partial v_j}{\partial
r_i}\right)^2, \nonumber
\end{eqnarray}
which is indeed proportional with $\nu X^2$ and
hence with biomass decay $F_{BD}=d_C C_C$ (which in the steady state equals the consumer growth $q_{CR}F_{BC}$).
\end{itemize}

To summarize, we have demonstrated a unique example of a correspondence between a
biological and a physical system. The dynamical equations are equivalent and a dictionary was given between the different quantities. Also the physical interpretations (in terms of energetic pathways) of the different terms in the dynamical equations were proven to be analogous. This correspondence allowed us to calculate an analytical expression for the heat transport in the steady state of the RB system. As Fig. \ref{energeticpathwaysfig} shows, a simple resource-consumer food web arises in the fluid system. In the next two sections, we will take
this analogy some steps further by expanding the fluid food web in two ways: First we will include competition at the first trophic level (the level of the consumers). Secondly, we will study longer food chains by including predation. In a sense, this approach allows us to use ecological concepts to extend the Lorenz dynamics in order to find new solutions (i.e. new convection patterns) for the fluid system.

\section{Competitive exclusion and fitness} \label{competitive exclusion}

In ecology, there is the important idea that species can mutate and evolve, leading to Darwinian competition between species. If we
describe competition in our ecosystem by taking
$n$ different consumer species with growth rates $g_{CRi}$, death rates
$d_{Ci}$ and yields $q_{CRi}$, with $i=1,\,...,n$, we can calculate
the stable steady state and it appears that the species with the highest
value of the competitive fitness
\begin{eqnarray}
f_i\equiv \frac{q_{CRi}g_{CRi}}{d_{Ci}}
\end{eqnarray}
survives, the others go extinct. This is a version of the famous
competitive exclusion principle (Armstrong and McGehee 1980).

As pointed out by Nicolis and Prigogine (1977),
in the fluid at the onset of convection, fluctuations in the form of
convection cells appear. These cells or rolls can have
different sizes, parametrized by the geometric factor $a$. Solving the Lorenz dynamics does not allow us to calculate the size of the convection rolls, because $a$ is treated as a constant parameter. But as the monoculture resource-consumer ecosystem can be generalized to a polyculture resource-consumers ecosystem, it is tempting to perform a translation in order to construct a generalization of the simplified Lorenz system. This adds a new element in the fluid systems: Rolls with different sizes (different $a_i$) will go into competition with each other.

With this generalization, one can can now ask which kind of convection cells are the most
fittest, which species of rolls will eventually survive. As the competitive exclusion principle states, the rolls with the highest fitness $f_i$ will survive, so the only thing we need to do is to translate the competitive fitness measure $f_i$ to the fluid system and write it as a function of the parameter $a_i$.
If we do the translation with the above
dictionary (\ref{C_R^0=T_H^0}-\ref{dC=nu}), we get the fluid fitness for rolls with parameter $a_i$:
\begin{eqnarray}
f_i\rightarrow\frac{ a_i^2 h^3 g\alpha}{(a_i^2+1)^3\pi^3\nu\chi}
\label{fitness}
\end{eqnarray}
Note that the geometric factor (\ref{G}) appears in the fitness. There is a trade-off between small and large sizes, and the fitness (\ref{fitness}) is maximal for rolls with parameter $a_i=1/\sqrt{2}$, and hence with width $h/a_i=\sqrt{2}h$. As was first shown by Rayleigh (1916) using a totally different line of reasoning, this is also
the experimentally verified size of the convection rolls at the onset of
convection. Furthermore, using this value for $a$ together with (\ref{C_Rcrit}) and the definition of the Rayleigh number (\ref{Ra}), we can calculate the critical Rayleigh number $Ra_c=27\pi^4/4$. This is indeed the correct value for the fluid system with free-free boundary conditions\footnote{This result is non-trivial, as the final words of appendix \ref{appendix XYZ} point out.}.

\section{Predating fluid motion}

A next
natural step to take is describing our ecosystem with the addition
of predators eating the consumers. This leads us to a more
speculative idea: Is there a possibility for 'predation' in fluid
systems? Let us first study the resource-consumer-predator ecosystem

The dynamical equations for the resource is the same as (\ref{dyn C_R}). For
the consumer and the predator the dynamics changes to
\begin{eqnarray}
\frac{d}{dt}C_C &=& q_{CR} g_{CR} (C_R-C_W^0/K_{eq}) C_C - g_{PC} C_C C_P - d_C C_C, \\
\frac{d}{dt}C_P &=& q_{PC} g_{PC} C_C C_P - d_P C_P.
\end{eqnarray}
There is now a second critical bifurcation point
\begin{eqnarray}
\Delta_{c2}^0&\equiv&
\frac{(\alpha_R+\kappa_{AC})}{\alpha_R}\frac{d_C}{q_{CR}g_{CR}}(1+\frac{d_P}{\alpha_R+\kappa_{AC}})
\label{C_R^II}
\end{eqnarray}
such that for values $\Delta^0\leq \Delta^0_{c2}$ we get the
previous solutions (\ref{C_R^*1}-\ref{C_C^*2}). For values higher
than this second critical concentration level, the former states
become unstable and the new stable state has a non-zero predator
concentration:
\begin{eqnarray}
C_R^* &=& \frac{q_{PC}g_{PC}\alpha_R \Delta^0}{q_{PC}g_{PC}(\alpha_R+\kappa_{AC})+g_{CR}d_P}+\frac{C_W^0}{K_{eq}}, \label{C_R predator}\\
C_C^* &=& \frac{d_P}{q_{PC}g_{PC}},\\
C_P^* &=& \frac{q_{CR}g_{CR}q_{PC}\alpha_R
\Delta^0}{q_{PC}g_{PC}(\alpha_R+\kappa_{AC})+g_{CR}d_P}-\frac{d_C}{g_{PC}}.
\end{eqnarray}

Moving to the convective fluid system, we have to study convection patterns that appear beyond a second bifurcation point. As shown above, there is a first bifurcation from conduction to straight
convection rolls. In the straight rolls situation, there was only
velocity in the x- and z-directions, leading to a non-zero kinetic
energy for these two directions. This $E_{kin,xz}$ was shown to be
related with the consumer concentration. But for certain systems
(depending on e.g. the Prandtl number), due to the
appearance of a velocity gradient in these rolls, there might be
changes in the surface tension leading to a new instability at a
second critical gradient level. This was experimentally as
well as numerically shown (Clever and Busse, 1987, Getling 1998). At this second bifurcation a new
pattern arises, from straight rolls to zig-zag rolls or rolls with
travelling waves in the direction of its rotation axis (the
$y$-direction). In these new patterns, there is also a non-zero
velocity component $v_y$ in the y-direction, leading to a non-zero kinetic
energy $E_{kin,y}$.

This allows us to propose a conjecture. The Lorenz system was derived by simplifying the Navier-Stokes equations in the Boussinesq approximation. By taking the lowest modes in an expansion, and performing an approximation, the Lorenz system was derived in order to study straight convection rolls. The wavy pattern could not be studied with the Lorenz dynamics. The conjecture states that by including another mode, a new variable that describes the motion in the $y$-direction, a new set of dynamical equations can be given (after performing some approximations to guarantee that the equations are autonomous), and this set of equations can be translated into the dynamics of a resource-consumer-predator ecosystem.

More specifically, the hypothesis that one can make is that the predator concentration
is proportional with the kinetic energy of $v_y$. The interpretation is that the waves are
behaving as predators feeding on the velocity gradient (or kinetic
energy) of the 'consumer prey' rolls, in a similar way as the
consumer prey rolls are feeding on the temperature gradient (heat
energy).

We did not prove this conjecture at the level of the dynamical equations, but
one will only give some (intuitive) arguments.

First, by looking at the advection term in
the heat equation (\ref{dyn T}), one can see that there is a
coupling between temperature and velocity, and it is this coupling
that was proven to be equivalent with the coupling of consumers with
the resource in the ecosystem dynamics. Now, by looking at the
advection term in the Navier-Stokes equation (\ref{dyn v}), one can
see that there is indeed a coupling between different velocity
components, so one might expect that this results in an equivalent
coupling between predators and consumers.

Second, our conjecture implies that the predator parameters are related to
the fluid parameters, in a similar way as in
(\ref{C_R^0=T_H^0}-\ref{dC=nu}). One might intuitively guess that
e.g. $d_P\sim \nu$. As can be seen in (\ref{C_R^II}), a term
$\frac{d_P}{\alpha_R+\kappa_{AC}}$ appears. As this is the ratio of viscosity over conductivity, this term is proportional with the Prandtl number (\ref{Prandtl}). The prefactor is dependent on the geometric factor which now includes the wavelength. As shown in e.g. Busse (1978), the second critical gradient level increases when
the Prandtl number increases. This is consistent with the increasing behavior observed in
(\ref{C_R^II}).

\begin{figure}[!ht]
\centering
\includegraphics[scale=0.40]{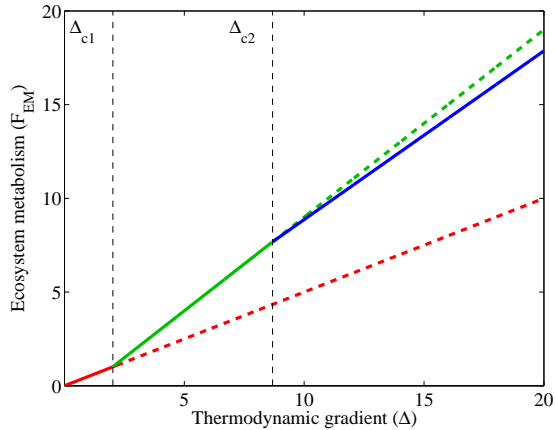}
\caption{The total steady state ecosystem metabolism $F_{EM}^*$ in function of
the external gradient $\Delta^0$, for the predator ecosystem. One can clearly see the existence
of two critical bifurcation points. Beyond the second bifurcation,
the rate in the predator state (blue) is lower than the consumer
state (dashed green).} \label{ratepredatorfig}
\end{figure}

A third test for this 'predator - kinetic energy' hypothesis is performed by
looking at the thermodynamical level. If our conjecture is correct,
the total steady state heat transport should be related with the
ecosystem metabolism in the steady state, as in
(\ref{energydissipation}). Using (\ref{C_R predator}), the latter can be easily calculated and
is presented in Fig. \ref{ratepredatorfig}. We see that for input
concentrations above the second bifurcation point, when predation is
possible, the stable predator state has always a \emph{lower}
ecosystem metabolism rate than the unstable consumer-only state.
Looking for example at the behavior of the Nusselt number (the dimensionless number which is proportional with the total heat transport) in the fluid system (Fig. 6 in Clever and Busse, 1987), we can see
that for all studied parameters, the heat transport in the wavy roll
state is indeed always lower than in the straight roll state.

Hopefully, one can rigorously proof this correspondence between ecological and fluidal predation. This would allow us to require more analytical expressions instead of using numerical simulations (Clever and Busse, 1987). Furthermore, if
this would be possible, we get a new parameter, the wavelength of
the zig-zag or wavy pattern which might be related with the
parameters $q_{PC}$, $g_{PC}$ and $d_P$. Perhaps it is possible to
derive the experimentally correct wavelength (see Pomeau and Manneville, 1980) again from competition and fitness at
the predator level, because the competitive exclusion principle also
works at this level (Smith and Waltman, 1995).

\section{Conclusions and further discussions}

We have seen that one can simplify the dynamical equations of a
convective fluid system into a set of two ordinary differential
equations which look exactly the same as a simplified
resource-consumer ecosystem. Furthermore, there is not only a
mathematical correspondence in the structure of the equations, but
more remarkable, there is also a correspondence between physical
interpretations. This correspondence was then broadened to include competition and predation. With these extensions, we have proven or conjectured more connections.
\begin{itemize}
\item We were able to calculate the correct value of the size of convection cells with the help of biological competition and fitness.
\item we have translated quantities, processes, energetic pathways,... from fluid systems to ecological systems and vice versa,
\item and we have conjectured a possible explanation of the decrease in energy dissipation in the fluid system after a second bifurcation point as being related with the appearance of predative behavior.
\end{itemize}

\begin{table}[!ht]
\begin{center}
\begin{tabular}{|c||c|c|}
  \hline
  & ecosystem & fluid system\\
  \hline
  \hline
  gradient & $\Delta^0=C_R^0-\frac{C_W^0}{K_{eq}}$ & $h\beta/2$ \\
  \hline
  variables & $C_R$& $T_H$ \\
  & $C_C$ & $X^2/g\alpha h^3$ ($E_{kin,xz}$) \\
  & $C_P$& $E_{kin,y}$  \\
  \hline
  growth & $g_{CR}$ & $g\alpha ha^2\pi^2/(a^2+1)\chi$\\
  \hline
  yield & $q_{CR}$ & $8/\pi^4(1+a^2)$  \\
  \hline
  decay & $d_C$ & $2\pi^2(a^2+1)\nu/h^2$ \\
  \hline
  exchange & $\alpha_R$, $\kappa_{AC}$ & $2\pi^2\chi/h^2$\\
  \hline
  flux & $\alpha_R(C_R^0-C_R^*)$ & $\pi^2W^*/2h\rho_0 c_V$ \\
  \hline
  fitness & $q_{CR}g_{CR}/d_C$ & $g\alpha h^3a^2/\nu\chi\pi^3(a^2+1)^3$\\
  \hline
\end{tabular}
\end{center}
\caption{The correspondence}\label{table correspondence}
\end{table}

Table \ref{table correspondence} presents the dictionary of the correspondence which is a quantitative extension of the bare essentials given in table \ref{table gradient degradation}. There is always unstructured gradient degradation, but above a critical level of the gradient, ordered patterns or structures appear: living cells and convection cells. This striking analogy that we have found between two systems that are at first sight totally different, can be casted in the (more general but often vague) language of dissipative structures used by Prigogine and co-workers (Glansdorff and Prigogine, 1971, Nicolis and Prigogine, 1977). The structured patterns of the RB system are often putted forward as prime examples of dissipative structures. It is believed that life also behaves as a dissipative structure. Prigogine and co-workers performed quantitative studies of biological systems, but these were mostly restricted to the subcellular level. Schneider and Kay (1994) took the correspondence further to the ecosystem level, but their discussion was only qualitative, using often vague words. Furthermore, in recent decades a new 'maximum entropy production' (MaxEP) school emerged (Kleidon, 2004; Kleidon and Lorenz 2005; Martiouchev and Seleznev, 2006) where it is believed that complex processes, including life, tend to maximize the entropy production. With our work, we extended the program initiated by Prigogine, Schneider, Kay and others by studying quantitatively the (thermodynamic) properties of dissipative structures at the ecosystem level. This lead to a more exact formulation of the correspondence, but also a new feature appeared, something which was not studied by Schneider et al.: the appearance of 'predative dissipative structures' after a second bifurcation. Not only is Fig. \ref{ratepredatorfig} an extension of Fig. \ref{ratesconsumerfig} (which was shown to be equivalent with Fig. 2a in Schneider and Kay, 1994), it also shows that the total gradient degradation (by heat transport or ecosystem metabolism) of the consumer-predator state is \emph{lower} than the corresponding unstable consumer state. As the gradient dissipation is proportional with the total entropy production, this might be a criticism on the basic hypotheses of the MaxEP-school. For high
thermodynamic gradients the energy dissipation of the state with
'second level' predative dissipative structures is \emph{lower} than
the state with only 'first level' dissipative structures. The predative dissipative structures
make the system less efficient in degrading the thermodynamic
gradient.

There are many new questions about the fluid- ecological system
analogy. We were able to determine the size of convection rolls by translating the Darwinian view of evolution and natural selection to the fluid system. Can this be generalized, i.e. is the difficult pattern selection problem in fluid systems (Getling, 1998) analogous to the difficult problem of
evolution and natural selection in ecology? More specifically: What about 'predatory' pattern selection (e.g. the selection of the wavelength of the wavy rolls)?  What about turbulent fluid states,
longer trophic chains (top-predation), 'fluidal niches and food webs', evolution
at different time scales, genetic information, velocity
correlations,...? Up till now, we were not yet able to derive new
non-trivial results, because solved problems were related with
solved ones, and unsolved with unsolved ones. We hope that besides
the esthetically pleasing results we have found, one is able to use
the analogy to find new solutions to important problems, both in ecology and fluid physics.

%\renewcommand{\thesection}{A}
  % redefine the command that creates the equation no.
  %\setcounter{equation}{0}  % reset counter 
\renewcommand{\theequation}{A-\arabic{equation}}
  % redefine the command that creates the equation no.
  \setcounter{equation}{0}  % reset counter 

\appendix

\section{The RCW ecosystem} \label{appendix RCW}

The resource-consumer-waste ecosystem consists of two environmental reservoirs, one for the resource and one for the waste. As an example, we can think of a chemotrophic ecosystem with glucose or methane as resource and $CO_2$ as waste product.
The resource
is supplied from the environmental reservoir at a fixed concentration
$C_R^0$ using a linear exchange mechanism with rate constant
$\alpha_R$ and flux
\begin{eqnarray}
F_R=\alpha_R(C_R^0-C_R).
\end{eqnarray}
$C_R(t)$ is the variable resource concentration in the ecosystem. The ecosystem metabolism is the total conversion (degradation) of resource into waste. In our chemotrophic ecosystem, this conversion is an oxidation proces. Table \ref{table ecosystemtransformations} shows the three metabolic transformations that occur within the
ecosystem, together with the kinetic expressions used.

\begin{table}[!ht]
\begin{center}
\begin{tabular}{|c|c|c|}
  \hline
  Abiotic conversion & $R \rightarrow W$ & $F_{AC} = \kappa_{AC}\left( C_R-\frac{C_W^0}{K_{eq}}\right)$ \\
  \hline
  Biomass synthesis & $R \,\,\,\, \rightarrow \,\,\,\,q_{CR}C$ \qquad & $F_{BC} = g_{CR}C_C\left( C_R-\frac{C_W^0}{K_{eq}}\right)$ \\
  and biotic conversion &  \qquad$ +(1-q_{CR})W$ & \\
  \hline
  Biomass decay & $C \rightarrow W$ & $F_{BD} = d_C C_C$\\
 \hline
\end{tabular}
\end{center}
\caption{Ecosystem transformations}\label{table ecosystemtransformations}
\end{table}

The abiotic conversion is a chemical reaction with equilibrium constant $K_{eq}$ and a constant abiotic conversion rate parameter $\kappa_{AC}$. The latter abiotic conversion rate is increased due to a parallel biotic conversion, described by a simple linear functional response with parameter $g_{CR}$. This biotic conversion has two parts: a fraction of the resource is used for consumer growth, the other part of the resource turns immediately into waste. From a thermodynamic perspective, the latter resource turnover is necessary to drive the growth process. This fractioning is described by the yield parameter $q_{CR}<1$: this is the growth efficiency which denotes the amount of resource required to build up one unit of biomass. The third metabolic transformation is the biotic decay (biomass turnover), represented by the rate constant $d_C$.

When the resource is turned into waste, the latter is emitted into the waste reservoir from the environment. The latter has a constant waste concentartion $C_W^0$ and the exchange flux can be described as
\begin{eqnarray}
F_W = \alpha_W(C_W^0-C_W).
\end{eqnarray}

Putting the two exchange fluxes and the three metabolic fluxes together, the complete dynamics for the resource concentration $C_R(t)$, the
consumer biomass concentration $C_C(t)$ and the waste concentration $C_W(t)$ now look like
\begin{eqnarray}
\frac{d}{dt}C_R &=& \alpha_R(C_R^0 - C_R) -(\kappa_{AC}+g_{CR}
C_C)(C_R-C_W/K_{eq}),\\
\frac{d}{dt}C_C &=& q_{CR} g_{CR} (C_R-C_W/K_{eq}) C_C - d_C C_C, \\
\frac{d}{dt}C_W &=& \alpha_W(C_W^0 - C_W)\nonumber\\ && + (\kappa_{AC} +
(1-q_{CR}) g_{CR} C_C)(C_R-C_W/K_{eq}) + d_C C_C. \label{dyn C_W}
\end{eqnarray}
This is the RCW model. Next, we have to simplify this model to the RC model, by assuming $\alpha_W$ to be very large. This means that the relaxation time of the waste exchange is negligibly small, and we get the condition that $C_W \approx C_W^0$, resulting into (\ref{dyn C_R}-\ref{dyn C_C}).

\renewcommand{\theequation}{B-\arabic{equation}}
  % redefine the command that creates the equation no.
  \setcounter{equation}{0}  % reset counter 

\section{The XYZ Lorenz system} \label{appendix XYZ}

In this appendix, we will give all approximations and a schematic derivation in order to arrive
at the Lorenz system for the Rayleigh-B\'enard convective fluid (see Berge and Pomeau, 1984 or Lorenz, 1963).

In order to present the field equations we will first list the
Boussinesq approximations (see e.g. Getling, 1998):
\begin{itemize}
\item There are no pressure terms in the energy balance equation.
\item The heat conduction coefficient $\chi$ and the kinetic viscosity $\nu$ are constants.
\item The local density field $\rho$ depends on the temperature as $\rho=\rho_0(1-\alpha(T-T_0))$ with $\rho_0$ and $T_0$ the constant reference density and temperature, $T$ the local temperature field, and $\alpha$ the constant thermal expansion coefficient.
\item The above dependence of the density on the temperature is taken into account only in the gravitational force term in the momentum balance equation. At other places in the equations, we will write the density as $\rho_0$.
\item The fluid is incompressible (except in the thermal expansion term): $\frac{d \rho}{dt}=0$, which results in an equality between heat capacities at constant pressure and volume: $c_p=c_v$, or it can be written in terms of the velocity field $\vec v$ as:
\begin{eqnarray}
\vec \nabla \cdot \vec v &=& 0, \label{div v}.
\end{eqnarray}
\item The local internal energy differential is $dU=c_pdT$.
\end{itemize}

With these approximations, the heat transport equation can be
derived from an energy balance equation, and looks like
\begin{eqnarray}
\frac{\partial T}{\partial t} &=& -(\vec v \cdot \vec \nabla) T
+\chi \Delta T. \label{dyn T}
\end{eqnarray}
The first term on the right hand side is the advective heat
transport term, and the second is the heat conduction term.

The equation for the velocity field is derived from the momentum
balance, and results into the Navier-Stokes equation. In the
Boussinesq approximation, this leads to
\begin{eqnarray}
\frac{\partial \vec v}{\partial t} &=& -(\vec v \cdot \vec
\nabla)\vec v -\frac{\vec \nabla p}{\rho_0}- g
\frac{\rho}{\rho_0}\vec 1_z +\nu \Delta \vec v, \label{dyn v}
\end{eqnarray}
with $p$ the pressure field, $g$ the gravitational acceleration and
$\vec 1_z$ the unit vector in the vertical z-direction. On the right
hand side we see respectively the advection term, the pressure
gradient term, the external gravitational force term and the viscous
diffusion term.

As a final step, in order to fully describe our system, we need
boundary conditions. The boundary condition for the temperature is
simply
\begin{eqnarray}
T(z=0)&=&T_H^0,\\
T(z=h)&=&T_C^0.
\end{eqnarray}
For the velocity, we have
\begin{eqnarray}
v_z(z=0)&=&0, \label{v_z(z=0)}\\
v_z(z=h)&=&0,
\end{eqnarray}
because there is no fluid flowing out of the
layer. This is not enough, and we need another condition on the
velocity. We will take free-free boundary conditions to make the
description of the solutions easier. This gives
\begin{eqnarray}
\frac{\partial
v_x}{\partial z}|_{z=0,h}=0.
\end{eqnarray}

In summary, we have five partial differential equations: Three from
the three velocity components, one from the incompressibility
condition and one from the temperature. Our five local variables are
the velocity, pressure and temperature fields. Lorenz made some
further assumptions in order to turn these five p.d.e.'s into three
o.d.e.'s with only three global variables.

Due to (\ref{div v}), one can write the velocity field as $\vec
v=\vec \nabla \times \vec\psi$, with $\vec\psi$ the streamfunction.
We know from experiment that at the onset of convection (near the
critical gradient), a convection roll pattern will arise
(Getling, 1998). Suppose that the axis of the rolls are along the
horizontal y-direction. Hence, there will be no $v_y$ component. The
simplest way to obtain this is by assuming $\psi_x=\psi_z=0$.

Next, we want to circumvent the pressure field. This can be done by
taking the curl of the velocity equation, resulting into:
\begin{eqnarray}
\frac{\partial \nabla^2 \psi_y}{\partial t}&=&\frac{\partial
\psi_y}{\partial z}\frac{\partial \nabla^2\psi_y}{\partial
x}-\frac{\partial \psi_y}{\partial x}\frac{\partial
\nabla^2\psi_y}{\partial
z}+\nu\nabla^2(\nabla^2\psi_y)+g\alpha\frac{\partial T}{\partial x}.
\label{curlPDEv}
\end{eqnarray}

As a final step, we will expand the temperature and $\psi_y$ fields
in Fourier modes, taking the boundary conditions into account, and
we will retain only three of these modes:
\begin{eqnarray}
\psi_y&=& X(t)\sin(\frac{\pi a x}{h})\sin(\frac{\pi z}{h}),\\
T&=&T_H^0-\beta z +Y(t)\cos(\frac{\pi a x}{h})\sin(\frac{\pi
z}{h})-Z(t)\sin(\frac{2 \pi z}{h}), \label{Temperature expansion}
\end{eqnarray}
with the width of the convection cell equal to $h/a$. In Fig.
\ref{YZandXfig} a physical interpretation is given to the variables
$X$, $Y$ and $Z$. Plugging these expressions into the above partial
differential equations (\ref{dyn T}) and (\ref{curlPDEv}), and
collecting the factors with the same spatial dependence, gives:
\begin{eqnarray}
\frac{d X}{d t}&=&-\frac{\nu (a^2+1)\pi^2}{h^2}X+\frac{g\alpha h a }{\pi (a^2+1)}Y, \label{dyn X Lorenz}\\
\frac{d Y}{d t}&=&-\frac{2\pi^2 a}{h^2}XZ\cos(\frac{2\pi z}{h})-\frac{\chi(a^2+1)\pi^2}{h^2}Y+\frac{\beta\pi a}{h}X, \label{dyn Y Lorenz with cos}\\
\frac{d Z}{d t}&=&\frac{a \pi^2}{2h^2}XY-\frac{\chi 4\pi^2}{h^2}Z \label{dyn Z Lorenz},
\end{eqnarray}
As can be seen, the system does not close because there is a
$\cos(2\pi z/h)$ term. A final approximation consists of
taking this cosine equal to one.

We finally arrive at the Lorenz
equations, which we will call the XYZ model. Next, we have to simplify this XYZ model to the XZ model, by assuming the pseudo steady state condition for $Y$ (i.e. taking $dY/dt=0$), resulting into (\ref{dyn X}-\ref{dyn Z}).

We conclude this appendix with an important remark. There are two important approximations for the XZ model. The first is the cancelation of the cosine factor in (\ref{dyn Y Lorenz with cos}). Therefore, solutions of the Lorenz system are not exact solutions of the complete fluid system in the Boussinesq approximation. In this sense, the result of section \ref{competitive exclusion} is not trivial, because we arrived at the correct answer whereas the underlying dynamics does not give exact solutions.  

Our second approximation is the pseudo steady state restriction. This means that the steady states of the XZ system are also steady states of the XYZ model (but as mentioned above, not necessarily of the complete fluid system). In this article we mostly restricted the discussion to the steady states of the XZ model, but one should be cautious to use this model to try to find correct transient solutions for the XYZ or the complete fluid systems. As an example, the XYZ system has chaotic solutions with unstable steady states, whereas these chaotic solutions are absent in the XZ model. The latter has always stable steady states.

%It is this friction that lies at the origin of the non-zero bifurcation point, as can be seen in (\ref{C_Rcrit}). Note also that the bifurcation point is highly sensitive on the functionality of the friction term. In our ecosystem model, this term was linear in $C_C$, but if it would have a higher order, e.g. $d_C C_C^2$, then the bifurcation point is at $\delta^0 = 0$ and the structures can survive arbitrarily close to equilibrium.

\end{document}